\documentclass[a4paper,12pt]{article}
\def\al{\alpha}
\def\be{\beta}
\def\ga{\gamma} 
\def\ep{\epsilon}
\def\lam{\lambda}
\def\Lam{\Lambda}

\def\calA{{\cal A}}   
  
 \def\calH{{\cal H}} 
  
 \def\calN{{\cal N}} 
\def\calP{{\cal P}}  \def\calR{{\cal R}}
\def\calS{{\cal S}}

\def\del        {  \partial  }
\def\half       {  {1\over 2}  }

\def\ie         {  {\it i.e.}      }

\def\comma          {\, ,}
\def\period         {\, .}
\def\lsim    {\lower .65ex \hbox{\ $\stackrel{<}{\sim}$\ } }
\def\gsim    {\lower .65ex \hbox{\ $\stackrel{>}{\sim}$\ } }
\def\com#1#2   { \left[#1, #2\right]} 
\def\acom#1#2  {\left\{ #1,#2\right\}}
\def\bra#1     {\langle #1 |}
\def\ket#1     {| #1 \rangle}
 
\def\vecii#1#2      {  \left(\begin{array}{c}#1\\#2\end{array}\right)  }
\def\veciii#1#2#3   {  \left(\begin{array}{c}#1\\#2\\#3\end{array}
                     \right)  }
\def\veciv#1#2#3#4  {  \left(\begin{array}{c}#1\\#2\\#3\\#4
                                 \end{array}\right)  }
\def\vecfv#1#2#3#4#5 {  \left(\begin{array}{c}#1\\#2\\#3\\#4\\#5
                                 \end{array}\right)  }

\def\matrixii#1#2#3#4            {  \left(\begin{array}{cc}#1&#2\\#3&#4
                                       \end{array}\right) }
\def\matrixiii#1#2#3#4#5#6#7#8#9 {  \left(\begin{array}{ccc}#1&#2&#3\\
                                     #4&#5&#6\\#7&#8&#9\end{array}
                               \right)  }
\def\mativ#1#2#3#4               {  \left(\begin{array}{cccc}
                                       #1\\#2\\#3\\#4\end{array}\right) }
\def\matv#1#2#3#4#5              {  \left(\begin{array}{ccccc}
                                     #1\\#2\\#3\\#4\\#5\end{array}
                              \right)  }
\def\eqabegin         {  \begin{eqnarray}  }
\def\eqaend           {  \end{eqnarray}  }
\def\nn               {  \nonumber  }
\def\bracetwo#1#2     {  \left\{ \begin{array}{l} #1 \\ #2 \end{array}
                         \right.  }
\def\bracetwocases#1#2#3#4  {   \left\{ \begin{array}{ll} #1 &
                                 \qquad #2 \\
                                 #3 & \qquad #4 \end{array} \right.  }
\def\bracebegin#1     {  \left\{ \begin{array}{#1}   }
\def\braceend         {  \end{array}\right.   }
\def\parn              {  \par\noindent }

\def\parmedskip        {  \par\medskip  }

\def\parag#1           {\paragraph{#1} \mbox{ }\parmedskip\noindent}
\def\msection#1      {  \begin{center} \section{#1} \end{center}   }
\def\nsection#1      {  \let\boldface\bf \def\bf{} \section{#1}
                           \let\bf\boldface   }
\def\mnsection#1     {  \begin{center} \nsection{#1} \end{center}  }
\def\capsection#1    {  \let\boldface\bf \def\bf{\sc} \section{#1}
                           \let\bf\boldface   }
\def\mcapsection#1   {  \begin{center} \capsection{#1} \end{center} }

\newcommand{\nullify}[1]{}

\def\papertitlepage{\baselineskip 3.5ex \thispagestyle{empty}}
\def\Title#1{\baselineskip 1cm \vspace{1.5cm}\begin{center}
 {\Large\bf #1} \end{center} 
\vspace{0.5cm}}
\def\Authors#1{\begin{center} {\it #1} \end{center}}
\def\Abstract{\vspace{1.0cm}\begin{center} {\large\bf Abstract} 
           \end{center} \par\bigskip}
\def\Komabanumber#1#2#3{\hfill \begin{minipage}{4.2cm} UT-Komaba #1
              \parn #2 
              \parn #3 \end{minipage}}
\renewcommand{\thefootnote}{\fnsymbol{footnote}}
\renewenvironment{thebibliography}{\pagebreak[3]\par\vspace{0.6em}
\begin{flushleft}{\large \bf References}\end{flushleft}
\vspace{-1.0em}

\begin{enumerate}\if@twocolumn\baselineskip=0.6em\itemsep -0.2em
\else\itemsep -0.2em\fi\labelsep 0.1em}{\end{enumerate}}

\usepackage{graphicx}
\usepackage{latexsym}
\usepackage{amsmath}
\usepackage{amssymb}

\renewenvironment{thebibliography}{\pagebreak[3]\par\vspace{0.6em}
\begin{flushleft}{\large \bf References}\end{flushleft}
\vspace{-1.0em}

\begin{enumerate}\if@twocolumn\baselineskip=0.6em\itemsep -0.2em
\else\itemsep -0.2em\fi\labelsep 0.1em}{\end{enumerate} }

\def\Btil{\tilde{B}}

\def\Itil{{\tilde{I}}}
\def\Jtil{{\tilde{J}}}
\def\Ktil{{\tilde{K}}}

\def\Rtil{\tilde{R}}
\def\Qtil{{\tilde{Q}}}

\def\deltatil{\tilde{\delta}}
\def\dtil{\tilde{d}}

\def\Bbar{{\bar{B}}}

\def\dbar{\bar{d}}

\def\Qbar{\bar{Q}}

\def\deltabar{\bar{\delta}}

\def\dhat{\hat{d}}

\def\lamhat{\hat{\lam}}
\def\Qhat{\hat{Q}}
\def\Qp{{Q'}}

\def\gcom#1#2{{\left[ #1, #2\right\}}}

\def\xp{x^+}

\def\psip{\psi^+}
\def\psim{\psi^-}

\def\lampinv{\lam_+^{-1}}
\def\lamp{\lam_+}
\def\lamI{\lam_\Itil}

\def\delx{\del x}
\def\delxp{\del x^+}
\def\delxm{\del x^-}
\def\thI{\theta_\Itil}
\def\thJ{\theta_\Jtil}
\def\thp{\theta_+}
\def\lamIJ{\lam_{IJ}}

\def\thIJ{\theta_{IJ}}
\def\thKL{\theta_{KL}}
\def\thMN{\theta_{MN}}

\def\Qp{Q_+}
\def\Qm{Q_-}

\def\Llam{L_{(\lam)}}

\def\QB{Q_B}
\def\QR{Q_{RNS}}
\def\QpR{Q'_{RNS}}
\def\QchR{\check{Q}_{RNS}}

\def\T{T}
\def\dch{\check{d}}
\def\Qch{\check{Q}}
\def\deltach{\check{\delta}}
\def\imag{i}

\def\XmT#1{x^-_{#1}}

\def\thIT#1{\theta_{#1}}

\def\bIT#1{b_{#1}}
\def\cIT#1{c_{#1}}
\def\thIJT#1{\theta_{#1}}

\def\lamIJT#1{\lam_{#1}}

\def\andeq{&=&}
\def\brkeq{\nn\\ &&\quad}
\def\npb#1{Nucl. Phys. {\bf B#1}}
\def\prd#1{Phys. Rev. {\bf D#1}}
\def\jhep#1{JHEP {\bf #1}}
\def\hepth#1{ hep-th/#1}
\def\plb#1{Phys. Lett. {\bf B#1}}

\begin{document}
\papertitlepage
\vspace*{0cm}
\Komabanumber{03-9}{hep-th/0305221} {May, 2003}
\Title{Operator Mapping between RNS and Extended Pure Spinor Formalisms
for Superstring} 
\Authors{{\sc Yuri Aisaka\footnote[2]{yuri@hep1.c.u-tokyo.ac.jp} 
 and Yoichi Kazama
\footnote[3]{kazama@hep1.c.u-tokyo.ac.jp}
\\ }
\vskip 3ex
 Institute of Physics, University of Tokyo, \\
 Komaba, Meguro-ku, Tokyo 153-8902 Japan \\
  }
\baselineskip .7cm

\numberwithin{equation}{section}
\numberwithin{figure}{section}
\numberwithin{table}{section}

\parskip=0.9ex

\Abstract

An explicit operator mapping in the form of a similarity transformation 
is constructed between the RNS formalism and an extension of  
the pure spinor formalism (to be called EPS formalism) recently proposed by 
the present authors. Due to the enlarged field space of the EPS
formalism, where the pure spinor constraints are removed,
the mapping is completely well-defined in contrast to the one given previously 
by Berkovits in the original pure spinor (PS) formalism. 
This map provides a direct demonstration of the equivalence of 
the cohomologies of the RNS and the EPS formalisms
and is expected to be useful for better understanding of 
various properties of the PS and EPS formalisms. 
Furthermore, the method of construction, 
which makes  systematic use of the nilpotency of the BRST charges, 
should find a variety of applications.

\newpage
\baselineskip 3.5ex
\section{Introduction}  
\renewcommand{\thefootnote}{\arabic{footnote}}
It has already been some time since a new formulation of a superstring 
 in which both the spacetime supersymmetry and the ten-dimensional Poincar\'e
 symmetry are manifest has been proposed by Berkovits \cite{Berk0001}, 
 following earlier attempts 
 \cite{Siegel86}\nocite{BVparticles,BVparticlesII,%
HSstring,TWstring,Hybrid}-\cite{U5form}. 
The central element of this so-called pure spinor (PS) 
formalism is the BRST-like charge 
 $\QB= \int [dz] \lam^\al d_\al$, where $d_\al$ is the spinor covariant 
 derivative and $\lam^\al$ is a bosonic chiral pure spinor 
\cite{Cartan,Nil86,Howe91}
satisfying the quadratic constraints 
$\lam^\al \ga^\mu_{\al\be} \lam^\be=0$. Under 
 these constraints, $\QB$ is nilpotent and its cohomology was shown to 
 reproduce the physical spectrum of a superstring \cite{Berk0006}. 
All the basic worldsheet fields in this formalism are free and form 
 a centerless conformal field theory (CFT). This allows one to 
 construct  $\QB$-invariant vertex operators 
 \cite{Berk0001,Berk0204} 
 and, together with 
 certain proposed rules, the scattering amplitudes 
 can be computed in a manifestly super-Poincar\'e covariant manner, 
 which agree with known 
results \cite{Berk0001,Berk0004,Trivedi:2002mf,Berk0104}.
Further developments and  applications of this formalism are 
 found in 
\cite{Berk0009}\nocite{Berk01050,Berk01051,%
Berk0112,Berk0201,Berk0203,Berk0205,%
Oda0109,Matone0206,stonybrook,Chesterman,Oda0302}-\cite{AK1}, and 
 a comprehensive review, up to a certain point,  is available
 in \cite{Berk0209}. 

Although a number of remarkable features have already been uncovered, 
 many challenges still remain for the PS formalism. The most demanding 
 is the understanding of the underlying fundamental action, its symmetry 
structures and quantization procedure. In order to achieve this goal, 
one needs to examine this formalism critically and try to gain as many 
hints as possible for the proper framework.  From this perspective, 
the non-linear constraints defining the very notion of pure spinor appear
to lead to some complications: Not only is it difficult to imagine that 
 a free quantized spinor with constraints emerge naturally 
in the future fundamental formalism but also the existence of 
these constraints presents a trouble in defining a proper inner product 
 structure, as pointed out in \cite{AK1}. 
Furthermore, as we shall discuss in more detail below, due to the 
 constrained field space one encounters a singular operation 
 in the process of relating the PS formalism to the conventional
 RNS formalism \cite{Berk0104}. 

Motivated by these considerations, in a recent work \cite{AK1} we have 
constructed an extension of the PS formalism, to be referred to 
 as EPS formalism,  where the PS constraints are 
 removed. As will be briefly reviewed in the next section, this is achieved 
 by an introduction of a minimum number (five) of fermionic ghost-antighost 
 pairs $(c_\Itil,b_I)_{I,\Itil=1\sim 5}$, 
which properly compensate the effects of 
the five components of $\lam^\al$ (and their conjugates) now freed from 
  constraints. It turned out that our formalism fits beautifully into 
 a mathematical scheme known as {\it homological perturbation theory} 
\cite{HenTb} and a genuine nilpotent BRST-like charge $\Qhat$, 
the cohomology of  which is guaranteed to be equivalent to that of $\QB$, 
was obtained. This scheme also provided  a powerful method of 
constructing the vertex operators, both integrated and unintegrated, 
which are the extensions of the ones in the PS formalism. Moreover, as an 
 important evidence of the advantage of the extended formalism, 
 we have been able to construct a remarkably simple composite 
 ``$b$-ghost'' field $B(z)$, which realizes the fundamental relation 
 $T(z) = \acom{\Qhat}{B(z)} $, where $T(z)$ is the Virasoro 
operator of the system. This has never been achieved in the 
PS formalism\footnote{Although we do not have a rigorous proof, an 
 analysis presented in Sec.~3.5 strongly indicates that 
 without  the extra degrees of freedom introduced in EPS, such a ``$b$-ghost''
 cannot be constructed. }.
 Another advantage of this formalism is 
 that the problem of defining a proper inner product can be solved 
 in the extended space without PS constraints \cite{AK1}. 

Besides advantages, we must mention an apparent disadvantage. Namely, 
due to the extra ghosts $(b_I, c_\Itil)$,  manifest ten-dimensional 
Lorentz covariance is broken down to $U(5)$ covariance. One need not, however, 
 regard this as a serious problem for two reasons. First, even in the 
 original PS formalism, in order to define the quantized fields properly, 
 one needs to solve the PS constraints and expresses the dependent components 
 of $\lam^\al$ in terms of independent components. This breaks the manifest 
 symmetry down to $U(5)$. Second, such a breakdown due to the ghosts
 is expected to be confined in the unphysical
 sector. Again the situation is very similar to that in the original
 PS formalism: As argued  by Berkovits, Lorentz-noncovariant effects 
in PS formalism can be decoupled from the physical quantities. 
Finally, we should mention that an alternative scheme of removing
 the pure spinor constraints with a finite number of ghosts
 has been developed in \cite{stonybrook}.
This formalism has the merit of retaining the Lorentz covariance
throughout but to get non-trivial cohomology one must impose
 an extra condition and this makes the formalism rather involved. 
Also, another scheme for the superparticle case has been proposed
 in \cite{Chesterman}.

One of the important remaining tasks for the EPS formalism is to clarify 
 how one can compute the scattering amplitudes using the vertex operators 
 constructed in \cite{AK1}. Just as in the PS formalism, one here encounters 
 a difficulty, in particular, 
 concerning the treatment of the zero modes. At the fundamental level, 
 this problem cannot be solved until one finds the underlying action 
 and derives the proper functional measure by studying 
how to gauge-fix various 
local symmetries. In the case of PS formalism, Berkovits circumvented 
 this process by ingeniously postulating a set of covariant rules which lead to
 the known results \cite{Berk0001,Berk0004,Trivedi:2002mf}. 
Further, the validity of these rules was supported by arguments relating
 PS to RNS \cite{Berk0104}. With the knowledge of the 
 underlying action still lacking, we must resort to similar means. 
In this paper, we shall construct, as a first step, 
 a precise operator  mapping between the EPS and the RNS formalisms 
entirely in the 
 form of a similarity transformation, which is  the most transparent way 
 to connect two theories.

 Before explaining our methods and results, 
we should briefly comment on the corresponding study in the PS formalism
  \cite{Berk0104}. In this work, by making  judicious identifications 
of the fields of the RNS formalism and those of the 
PS formalism, the extended BRST operator\footnote{$\eta_0$
 is the zero mode of the $\eta$ field appearing in the well-known 
 ``bosonization'' of the $\be$-$\ga$ bosonic ghosts \cite{FMS86}.}
 $\QpR\equiv \eta_0+\QR$ in the so-called large Hilbert space 
is expressed in terms of PS variables. Then, certain degrees of 
 freedom of the PS formalism which are missing in the RNS are added 
 in such a way to keep intact the nilpotency 
of the BRST charge as well as  the physical content of the theory.  Finally, 
 by a similarity transformation, the BRST charge so modified is 
 mapped to the one appropriate for PS \ie  to $\QB$. Although 
most of these  manipulations are rather natural, 
 the similarity transformation 
 employed in the last step contains a singular function and leads to 
 some difficulties, as discussed at some length in \cite{Berk0104}.
 This can be traced to the imposition of the non-linear PS constraints. 

Our method to be developed in this paper for connecting 
 the EPS and the RNS formalisms is rather different, and is 
intimately linked to the scheme of 
 homological perturbation theory. It enables us to construct
 in a systematic way a complete similarity transformation
 which maps the BRST-like charge $\Qhat$ of the EPS to the 
extended BRST charge $\QpR$ of the RNS, modulo 
 cohomologically trivial orthogonal operators. Due to the absence of the 
PS constraints, this mapping is entirely well-defined. We believe that 
 this powerful method has not been recognized before and should have 
many useful applications. 

Let us now give an outline of our procedures and results,
 which at the same time serves to 
indicate the organization of the paper. 
After a brief review of the PS and the EPS formalisms in Sec.~2,
 we begin Sec.~3 with a comparison of the 
 degrees of freedom of EPS, PS and RNS (Sec.~3.1.) This will make it evident 
 that EPS contains extra degrees of freedom compared to RNS 
in the form of two sets of BRST quartets, which we need to decouple. 
Since this task will be  somewhat involved, we shall first consider 
in Sec.~3.2 a simpler problem of constructing a 
similarity transformation that connects EPS to PS,  in order to illustrate 
our basic idea. Although such an equivalence was already proven in \cite{AK1}, 
 this provides an alternative more direct proof. After this warm-up, 
the decoupling of the first quartet is achieved in 
 Sec.~3.3 and that of the second quartet in Sec.~3.4, both 
 by means of similarity transformations. This brings the original 
 $\Qhat$ to an extremely simple operator, to be called $\Qbar$, plus 
 trivial nilpotent operators which are orthogonal to $\Qbar$. 
Partly as a check of the similarity transformation, we study in Sec.~3.5
 how the $B$-ghost is transformed. This analysis makes more transparent
 the difficulty of constructing $B$-ghost in the PS formalism 
defined in constrained field space. 
In Sec.~4 we turn our attention to the RNS side and construct, by an 
 analogous method,  a similarity transformation which drastically 
simplifies $\QpR$ down to an operator which will be denoted as 
 $\eta_0 + Q_0$. With the BRST charges on both sides reduced to simple forms, 
 it is now an easy matter to establish their relations. In Sec.~5.1, 
 we display the identification of fields given in \cite{Berk0104} in 
  appropriate forms, check that these rules produce correct conversion 
 of the energy-momentum tensors and show that in fact the operators 
$\Qbar$ and $\eta_0 + Q_0$ are identical. 
Finally in Sec.~5.2 we discuss the important 
 problem of the restriction of the proper Hilbert space necessary to 
 achieve the correct cohomology on both sides. 
This completes the 
explicit demonstration of the equivalence of EPS, PS, and RNS formalisms. 
Sec.~6 is devoted to a brief summary and discussions. 
\section{A Brief Review of PS and EPS Formalisms  }
In order to make this article reasonably self-contained and 
at the same time to explain our notations, let us begin 
 with a very brief review of the essential features of the PS and the EPS
formalisms.  
\subsection{PS Formalism}
The central idea of the pure spinor formalism 
\cite{Berk0001} is that the physical states of superstring 
can be described as the elements of the cohomology of a BRST-like 
 operator $\QB$ given by\footnote{For simplicity we will use 
 the notation $[dz] \equiv dz/(2\pi i)$ throughout.}
\begin{eqnarray}
\QB &=& \int [dz] \lam^\al(z) d_\al(z) \comma \label{defQ}
\end{eqnarray}
where $\lam^\al$ is a $16$-component bosonic chiral spinor 
 satisfying the  pure spinor constraints 
\begin{eqnarray}
\lam^\al \ga^\mu_{\al\be} \lam^\be &=& 0 \comma \label{psconst}
\end{eqnarray}
and $d_\al$ is the spinor covariant derivative given in our 
convention\footnote{Our conventions, including normalization, 
 of a number of quantities are slightly different
 from those often (but  not invariably) used by  Berkovits. }
by 
\begin{eqnarray}
d_\al &=& p_\al + i\del x_\mu (\ga^\mu\theta)_\al + \half 
(\ga^\mu\theta)_\al (\theta \ga_\mu \del \theta) \period \label{defdal}
\end{eqnarray}
$x^\mu$ and $\theta^\al$ are, respectively, the basic bosonic and 
ferminonic worldsheet fields describing a superstring, which transform 
 under the spacetime supersymmetry with global spinor parameter $\ep^\al$ 
as $\delta \theta^\al = \ep^\al$, 
$\delta x^\mu = i\ep \ga^\mu \theta$.
$x^\mu$ is self-conjugate and satisfies  $x^\mu(z) x^\nu(w) 
 = -\eta^{\mu\nu} \ln (z-w)$, while $p_\al$ serves as the conjugate 
 to $\theta^\al$ in the manner $\theta^\al(z) p_\be(w) 
 = \delta^\al_\be (z-w)^{-1}$.
 $\theta^\al$ and $p_\al$ carry 
 conformal weights 0 and 1 respectively. With such free field operator 
 product expansions (OPE's), 
$d_\al$ satisfies the following OPE with itself, 
\begin{eqnarray}
d_\al (z) d_\be(w) &= & {2i \ga^\mu_{\al\be} \Pi_\mu(w) 
\over z-w} \comma \label{ddope}
\end{eqnarray}
where $\Pi_\mu$ is the basic superinvariant combination 
\begin{eqnarray}
\Pi_\mu &=& \del x_\mu -i\theta \ga_\mu \del \theta \period 
\label{defPi}
\end{eqnarray}
Then, due to the pure spinor constraints (\ref{psconst}), $\QB$ is easily 
 found to be nilpotent and the  constrained cohomology of $\QB$ 
 can be defined. The basic superinvariants $d_\al, \Pi^\mu$ and 
$\del\theta^\al$  form the closed algebra 
\begin{eqnarray}
d_\al(z) d_\be(w) &=& {2i\ga^\mu_{\al\be} \Pi_\mu (w)\over z-w} 
\comma \label{dd2}\\
d_\al(z) \Pi^\mu(w) &=& {-2i(\ga^\mu \del\theta)_\al(w) \over z-w} 
\comma \label{dPi2}\\
\Pi^\mu(z) \Pi^\nu(w) &=& -{\eta^{\mu\nu} \over (z-w)^2} 
\comma \label{PiPi2}\\
d_\al(z) \del \theta^\be(w) &=& {\delta^\be_\al \over (z-w)^2} 
\comma \label{ddelth2}
\end{eqnarray}
which has  central charges and hence is essentially of {\it second class}. 

Although eventually the rules for computing the scattering 
 amplitudes are formulated in a Lorentz covariant manner, proper
quantization of the pure spinor $\lam$ can only be performed  
by solving the PS constraints (\ref{psconst}), which inevitably 
 breaks covariance in  intermediate steps.
 One convenient scheme is the so-called $U(5)$ formalism\footnote{
Although our treatment applies equally well to $SO(9,1)$ and $SO(10)$ 
 groups, we shall use the terminology appropriate for $SO(10)$,
 which contains $U(5)$ as a subgroup. Details 
 of our conventions for $U(5)$ parametrization are found in 
 the Appendix A of \cite{AK1}.}, in which 
a chiral and an anti-chiral spinors $\lam^\al$ and $\chi_\al$, respectively, 
are decomposed in the following way
\begin{eqnarray}
\lam^\al &=& (\lam_+, \lam_{IJ}, \lam_\Itil) \sim  (1,10,\overline{5}) \comma
\label{lamal}  \\
\chi_\al &=& (\chi_-, \chi_{\Itil\Jtil}, \chi_I) 
\sim (1,\overline{10}, 5)\comma 
 \qquad (I,J, \Itil,\Jtil = 1 \sim 5)\comma   \label{chial}
\end{eqnarray}
where we have indicated how they transform under $U(5)$, with a 
 tilde on the $\bar{5}$ indices. On the other hand, a Lorentz vector 
 $u^\mu$ is split into $5+\overline{5}$ of $U(5)$ as 
\begin{eqnarray}
u^\mu &=& 2 (e^{+\mu}_I u^-_{\Itil} + e^{-\mu}_\Itil u^+_I ) \comma 
\end{eqnarray}
where the projectors $e^{\pm \mu}_I$, defined by 
 $e^{\pm \mu}_I \equiv \half (\delta_{\mu,2I-1} \pm i \delta_{\mu, 2I})$,
 enjoy the properties
\begin{eqnarray}
&& e^{\pm \mu}_I e^{\pm \mu}_J = 0 \comma \qquad 
  e^{\pm \mu}_I e^{\mp \mu}_J = \half \delta_{IJ} \comma \\
&&  e^{+\mu}_I e^{-\nu}_\Itil +e^{-\mu}_\Itil e^{+\nu}_I
 = \half \delta^{\mu\nu} \period
\end{eqnarray}
In this scheme the 
 pure spinor constraints reduce to 
 5 independent conditions\footnote{Here and hereafter, for simplicity 
 of notation, we shall denote $\ep_{\Itil\Jtil\tilde{K}
\tilde{L}\tilde{M}}$ as  $\ep_{IJKLM}$.}
\begin{eqnarray}
\Phi_\Itil \equiv  \lam_+\lam_\Itil -{1\over 8}
\ep_{IJKLM}\lam_{JK}
 \lam_{KL} =0  \comma  \label{defPhi} 
\end{eqnarray}
and hence $\lam_\Itil$'s are solved in terms of $\lamp$ and $\lam_{IJ}$. 
Therefore the number of independent components of a pure spinor is 
 11 and together with all the other fields (including the conjugates 
 to the independent components of $\lam$) the entire system constitutes 
 a free CFT with vanshing central charge. 

The fact that the constrained cohomology of $\QB$ is in one to one 
 correspondence with the light-cone degrees of freedom of superstring 
 was shown in \cite{Berk0006} using the $SO(8)$ parametrization of a
 pure spinor. Besides being non-covariant, 
this parametrization contains redundancy and an infinite
 number of supplimentary ghosts had to be introduced. Nonetheless, 
 subsequently the Lorentz invariance of the cohomology 
was demonstrated in \cite{Berk01051}. 

The great advantage of this formalism is that one can compute the 
 scattering amplitudes in a manifestly super-Poincar\'e covariant
 manner. For the massless modes, 
the physical unintegrated vertex operator is given by a simple form 
\begin{eqnarray}
U = \lam^\al A_\al(x,\theta) \comma \label{unintvert} 
\end{eqnarray}
where $A_\al$ is a spinor superfield satisfying the ``on-shell'' condition
$(\ga^{\mu_1\mu_2 \ldots \mu_5})^{\al\be} D_\al A_\be = 0$ 
with $D_\al = {\del\over \del\theta^\al} -i(\ga^\mu\theta)_\al
{\del \over \del x^\mu} $. 
Then, with the pure spinor constraints, one easily verifies $\QB U=0$ 
and moreover finds that $\delta U =\QB\Lam$ 
 represents the gauge transformation
 of $A_\al$. Its integrated counterpart $\int[dz]V(z)$, needed 
 for calculation of $n$-point amplitudes with $n\ge 4$, is 
 characterized by $\QB V =\del U$ and was constructed 
to be of the form \cite{Berk0001,Siegel86}
\begin{eqnarray}
V = \del\theta^\al A_\al + \Pi^\mu B_\mu + d_\al W^\al  
 + \half \Llam^{\mu\nu} F_{\mu\nu} \period \label{intvert}
\end{eqnarray}
Here, $B_\mu =(i/16) \ga^{\al\be}_\mu D_\al A_\be$ is the gauge superfield, 
 $W^\al =(i/20) (\ga^\mu)^{\al\be}(D_\be B_\mu-\del_\mu A_\be)$ is 
 the gaugino superfield,  $F_{\mu\nu}=\del_\mu B_\nu -\del_\nu B_\mu$ 
 is the field strength superfield and $\Llam^{\mu\nu}$ is the Lorentz generator
 for the pure spinor sector. 

With these vertex operators, the scattering amplitude is expressed 
 as 
$\calA =\langle U_1(z_1)U_2(z_2)$ \allowbreak $ U_3(z_3) 
\int [dz_4] V_4(z_4) \cdots \int [dz_N] V_N(z_N) \rangle$,
 and  can be 
 computed in a covariant manner with certain rules assumed for
 the integration over the zero modes of $\lam^\al$ and $\theta^\al$. 
The proposed prescription enjoys a number of required properties and 
 leads to results which agree with those obtained in the RNS formalism
\cite{Berk0001,Berk0004,Trivedi:2002mf,Berk0104}.

\subsection{EPS Formalism}
Although the PS formalism briefly reviewd above has a number of 
 remarkable features, for the reasons stated in the introduction, 
 it is desirable to remove the PS constraints by extending 
 the field space. Such an extension was achieved in a minimal manner 
 in \cite{AK1}. Skipping all the details, we give below the essence of the 
 formalism. 

Instead of the basic superinvariants forming the essentially 
 second class algebra (\ref{dd2}) 
 $\sim$ (\ref{ddelth2}), we introduce the four types of composite 
 operators
\begin{eqnarray}
j  &=& \lam^\al d_\al \comma \label{defj} \\
\calP_I &=& \calN^\mu_I \Pi_\mu \comma \label{defcalP} \\
\calR_{IJ} &=& 2i\lampinv \calN^\mu_I (\ga_\mu \del \theta)_J \comma 
\label{defcalR} \\
\calS_{IJ} &=& -(\del \calN^\mu_I) \calN^\mu_J \comma \label{defcalS} 
\end{eqnarray}
where $\calN^\mu_I$ are a set of five Lorentz vectors which are 
 {\it null},  \ie $\calN^\mu_I \calN^\mu_J =0$, defined by 
\begin{eqnarray}
\calN^\mu_I \equiv -4(e^{+\mu}_I -\lampinv \lam_{IJ} e^{-\mu}_\Jtil ) 
\period \label{defcalN}
\end{eqnarray}
Note that $j$ is the BRST-like current of Berkovits {\it now without 
 PS constraints}. 
The virtue of this set of operators is that they form a closed algebra 
 which is of {\it first class}, namely without any central charges. 
This  allows one to build a BRST-like nilpotent charge $\Qhat$ 
associated to this algebra. Introducing five 
sets of fermionic ghost-anti-ghost pairs $(c_\Itil, b_I)$ carrying 
 conformal weights $(0,1)$ with 
the OPE
\begin{eqnarray}
 c_\Itil(z) b_J(w) = {\delta_{IJ} \over z-w} 
\comma \label{bIcIope}
\end{eqnarray}
and making use of the powerful scheme known as homological perturbation
 theory \cite{HenTb}, $\Qhat$ is constructed as 
\begin{eqnarray}
\Qhat &=& \delta + Q + d_1 + d_2  \comma  \label{Qhat}
\end{eqnarray}
where 
\begin{eqnarray}
\delta &=& -i\int [dz] b_I \Phi_\Itil \comma \qquad 
Q = \int [dz] j \comma \label{deltaQ}\\
d_1 &=& \int[dz] c_\Itil \calP_I \comma \qquad 
d_2 = -{i\over 2}\int[dz] c_\Itil c_\Jtil \calR_{IJ} \period \label{d1d2}
\end{eqnarray}
The operators $(\delta, Q, d_1, d_2)$ carry degrees $(-1,0,1,2)$ 
under the grading  $\deg (c_\Itil) = 1$, $\deg (b_I) =-1$, 
$\deg (\mbox{rest})=0$ and the nilpotency of $\Qhat$ follows 
 from the first class algebra mentioned above. 

The crucial point of this construction is that by the main theorem of 
 homological perturbation the cohomology of $\Qhat$ is guaranteed to 
 be equivalent to that of $Q$ with the constraint $\delta=0$, \ie with 
 $\Phi_\Itil=0$, which are nothing but the PS constraints (\ref{defPhi}).
 Moreover,  the underlying logic of this proof can be adapted to construct 
the massless vertex operators, both unintegrated and integrated, which 
 are the generalization of the ones shown in 
(\ref{unintvert}) and (\ref{intvert}) for the PS formalism. 

To conclude this  brief review, let us summarize the basic fields 
of the EPS formalism, their OPE's, the energy-momentum tensor 
$T_{EPS}(z)$ and 
 the $B$-ghost field that realizes the important relation 
 $\acom{\Qhat}{B(z)} = T_{EPS}(z)$. Apart from the $(c_\Itil, b_I)$ ghosts 
 given in (\ref{bIcIope}), the basic fields are the conjugate pairs 
$(\theta^\al, p_\al)$, $(\lam^\al, \omega_\al)$, both of which carry 
 conformal weights $(0,1)$,  and the string coordinate
 $x^\mu$. Non-vanishing OPE's among them are
\begin{eqnarray}
\theta^\al(z) p_\be(w) &=& {\delta^\al_\be \over z-w}\comma 
 \quad \lam^\al(z) \omega_\be(w) = {\delta^\al_\be \over z-w} 
\comma \quad x^\mu(z) x^\nu(w) = -\eta^{\mu\nu} \ln(z-w) \comma \nn\\
\end{eqnarray}
which in $U(5)$ notations read
\begin{eqnarray}
\thp(z) p_-(w) &=& {-1\over z-w}\comma \quad \thI(z) p_J (w) = {-\delta_{IJ} 
 \over z-w}\comma \quad \thIJ(z) p_{\tilde{K}\tilde{L}}(w) 
 = {\delta^{KL}_{IJ} \over z-w} \comma \\
\lamp(z) \omega_-(w) &=& {-1\over z-w}\comma \quad \lamI(z) \omega_J (w)
 = {-\delta_{IJ} 
 \over z-w}\comma \quad \lamIJ(z) \omega_{\tilde{K}\tilde{L}}(w) 
 = {\delta^{KL}_{IJ} \over z-w} \comma \\
x^+_I(z) x^-_\Jtil(w) &=& -\half \delta_{IJ} \ln (z-w) \comma
\end{eqnarray}
where $\delta^{KL}_{IJ} \equiv \delta^K_I \delta^L_J - \delta^K_J \delta^L_I$.
The energy-momentum tensor is of the form 
\begin{eqnarray}
T_{EPS} &=& -\half \del x^\mu \del x_\mu -p_\al \del\theta^\al 
 -\omega_\al \del \lam^\al -b_I \del c_\Itil \nn\\
&=& -2\del x^+_I \del x^-_\Itil + p_- \del \thp + p_I \del \thI 
-\half p_{\Itil \Jtil} \del \thIJ \nn\\
&& \qquad + \omega_- \del \lamp + \omega_I \del \lamI 
-\half \omega_{\Itil \Jtil}
 \del \lamIJ -b_I \del c_\Itil \comma \label{TEPS}
\end{eqnarray}
with the total central charge vanishing. Finally, the $B$-ghost field is 
 given by 
\begin{eqnarray}
B &=& -\omega_\al \del \theta^\al + \half b_I \Pi^-_\Itil \period
\label{Bghost}
\end{eqnarray}
\section{Similarity Transformation in EPS }
\subsection{Comparison of the field-content of EPS and RNS}
As stated in the introduction, our aim in this work is to find a precise
 operator mapping between the EPS and the RNS formalisms. To do so, 
 we must first compare and clarify the field-content of these formulations. 
In the $U(5)$ notation, this is given in the following table, where 
 in parentheses contributions to the central charge are indicated:
\begin{eqnarray}
&& \mbox{EPS} \qquad x^\pm_I\quad \vecii{\theta_\Itil}{p_I} 
\quad \vecii{\theta_+}{p_-} \quad \vecii{\lam_+}{\omega_-}  \quad 
 \vecii{\lam_{IJ}}{\omega_{\Itil\Jtil}} \quad 
 \vecii{\theta_{IJ}}{p_{\Itil\Jtil}}
\ \  \vecii{\lam_\Itil}{\omega_I} \quad \vecii{c_\Itil}{b_I} \quad 
   \nn\\
&& \hspace{1.5 cm} (10) \quad (-10) \qquad\ (-2) \qquad \quad (2) 
 \hspace{1.4cm}  (20) \ \qquad \ (-20) \qquad \, (10) \qquad \ (-10) \nn\\
&& \nn\\
&& \mbox{RNS}\qquad x^\pm_I\quad \psi^\pm_I\quad  \vecii{ b}{c}
\quad \vecii{\beta}{\gamma}  \nn\\
&&  \hspace{1.5cm} (10) \quad (5) \quad (-26) \qquad (11)  \nn
\end{eqnarray}
On the RNS side, $\psi^\pm_I =e^{\pm\mu}_I \psi_\mu$ are the matter 
 fermions and $(b,c)$ and $(\be,\ga)$ are the familiar fermionic and bosonic 
 ghosts. 
By counting the number of bosonic and fermionic fields, one sees that, 
compared to the RNS, the EPS formalism contains extra degrees 
 of freedom forming two ``quartets''\footnote{The precise context in which 
 they form quartets will be explained later.}
 $(\lam_{IJ},\omega_{\Itil\Jtil}, \theta_{IJ}, p_{\Itil\Jtil})$ and $(\lamI, \omega_I, c_\Itil, b_I)$. (In the case of 
 the original PS formalism, the second quartet is absent.)  Therefore it is 
 clear that to connect EPS to RNS, one must decouple these quartets in an 
 appropriate way. This will be done in subsections 3.3 and 3.4. 

For the rest of the work, it will be convenient to use the standard 
``bosonized'' representations for the $\be$-$\ga$ ghosts 
 \cite{FMS86}. Namely, we write them as 
\begin{eqnarray}
\be &=& \del \xi e^{-\phi}\comma \qquad \ga=e^\phi \eta \comma \\
\xi &=& e^\chi \comma \qquad \eta = e^{-\chi} \comma \label{bosxieta}
\end{eqnarray}
where $(\xi,\eta)$ are fermionic ghosts with dimensions $(0,1)$ and 
$\phi$ and $\chi$ are chiral bosons satisfying the OPE
\begin{eqnarray}
\phi(z)\phi(w) &=& -\ln(z-w)\comma \qquad \chi(z)\chi(w) = \ln (z-w) 
\period
\end{eqnarray}
Also,  for some purposes bosonization of the  $b$-$c$ ghosts as well as 
the matter fermions $\psi^\pm_I$ will be useful
 as well:
\begin{eqnarray}
c &=& e^{\sigma}\comma \qquad b=e^{-\sigma} \comma 
\qquad \sigma(z) \sigma(w) = \ln (z-w) \comma \label{boscb}\\
\psip_I &=& {1\over \sqrt{2}} e^{-H_I}\comma \qquad 
\psim_\Itil =  {1\over \sqrt{2}} e^{H_I} \comma \qquad 
H_I(z) H_J(w) = \delta_{IJ} \ln(z-w) \label{bospsiI}\period
\end{eqnarray}
The sum of $H_I$ bosons will be denoted by $H \equiv \sum_I H_I$.

Now it is well-known \cite{FMS86} that the RNS string can be formulated 
either in the small Hilbert space $\calH_S$ without $\xi_0$, \ie the 
 zero mode of $\xi$,  or in  the large Hilbert space $\calH_L$ including
 $\xi_0$. Since the BRST-like charge $\Qhat$ for the EPS formalism 
 contains zero modes of all the relevant fields, one expects that,
 after the decoupling of the quartets,  EPS is connected to the RNS 
 formulated in the large Hilbert space.
 This will be elaborated further in sections 4 and 5. 
\subsection{Equivalence of EPS and PS by a similarity transformation}
Since the construction of the similarity transformation 
which decouples the two quartets described above is, as we shall shortly see, 
 somewhat involved, it is instructive to begin with a similar 
 but much simpler task of proving the equivalence of EPS and PS 
 formalisms by the method of similarity transformation, in order 
 to illustrate the basic idea and logic. This equivalence
 was already proven in our previous paper by the machinary of homological 
perturbation theory, and hence the following will serve as the second
 (and  more  direct) proof. 

The goal is to relate the BRST-like charges $\Qhat$ and $Q_B$, for EPS and 
PS formalisms respectively, by a similarity transformation. 
With PS constraints imposed, $Q_B$ can be written as 
\begin{eqnarray}
Q_B &=& \int[dz] \lamhat^\al d_\al \comma 
\end{eqnarray}
where  $\lamhat^\al$ is the pure spinor for which $\lam_{\Itil}$ components
 are replaced by $(1/8)\lampinv \ep_{IJKLM} \lam_{JK} \lam_{LM}$.
On the other hand, recalling the form of $\Qhat$ 
 given in (\ref{Qhat}) $\sim$ (\ref{d1d2}), 
its degree 0 component $Q$ is given by  $Q=
\int[dz] \lam^\al d_\al$ without any constraints on $\lam^\al$. Thus, 
evidently  $Q$ and $Q_B$ are related by 
\begin{eqnarray}
Q &=& Q_B + \Qbar\comma \qquad \Qbar \equiv -\int[dz] 
\lampinv \Phi_\Itil d_I \period
\end{eqnarray}
To go from $\Qhat$ to $Q_B$, we must obviously remove $\Qbar$. 
To this end, note that $\Qbar$ is linear in the PS constraint $\Phi_I$, 
and hence we should be able to write it as\footnote{Here and hereafter,
 a product $AB$ of two integrated operators
will always signify the operator product in the sense of conformal field 
 theory and hence is equal to the graded commutator $\gcom{A}{B} $. 
In this notation, the graded Jacobi identity reads 
$ABC \equiv A(BC) = (AB)C \pm B(AC)$.}
\begin{eqnarray}
\Qbar &=& \delta R_1  \comma 
\end{eqnarray}
where $R_1$ is an integrated operator 
of degree 1. 
 Such an operator is easily found and is given by 
\begin{eqnarray}
R_1 &=& -i \int [dz]  \lampinv c_I d_I \period
\end{eqnarray}
This suggests that we should use this $R_1$ as the exponent of the similarity 
 transformation, namely  $e^{R_1}\Qhat e^{-R_1}$. The relevant 
 calculations are easily performed with the aid of the OPE's between 
 the $U(5)$ components of $d_\al$ and $\Pi^\mu$, which follow 
 from (\ref{dd2}) $\sim$ (\ref{ddelth2}).  After some algebra, we find 
\begin{eqnarray}
R_1 R_1 \delta  &=& 0 \comma \qquad 
R_1 Q = -d_1 \comma \qquad R_1 d_1 = -2d_2 \comma \qquad 
 R_1 d_2=0 \period
\end{eqnarray}
This means that under the similarity transformation each part of $\Qhat$ gets
 transformed as 
\begin{eqnarray}
e^{R_1} \delta e^{-R_1} &=& \delta -\Qbar\comma \\
e^{R_1} Q e^{-R_1} &=& Q-d_1+d_2 \comma \\
e^{R_1} d_1 e^{-R_1} &=& d_1-2d_2 \comma \\
e^{R_1} d_2 e^{-R_1} &=& d_2 \period
\end{eqnarray}
Adding up, we get a remarkably simple (and expected) result:
\begin{eqnarray}
e^{R_1} \Qhat e^{-R_1} &=& \delta +Q_B \period
\end{eqnarray}
Since $\delta$ and $Q_B$ are nilpotent and anticommute with each other, 
 the main theorem  of homological perturbation theory 
tells us that the cohomology of 
$\delta +Q_B$ is the same as that of $Q_B$ with $\delta$ set to zero, 
 \ie with the PS constraint $\Phi_\Itil=0$. This proves in a direct way 
 the equivalence of $\Qhat$-cohomology and Berkovits' cohomology. 
\subsection{Decoupling of the first quartet}
We now launch upon the task of decoupling the quartets by a judicious 
similarity transformation. 

Consider first the decoupling of 
$(\lam_{IJ},\omega_{\Itil\Jtil}, \theta_{IJ}, p_{\Itil\Jtil})$. To this end, 
 we shall make use of a refined filtration used previously 
by Berkovits \cite{Berk0104}. Namely, we shall assign  non-vanishing 
 degrees to the fields we wish to separate in the following way:
\begin{eqnarray}
&& \deg(p_{\Itil\Jtil})= -2\comma  \qquad \deg(\thIJ)= +2 \comma 
\label{deg1pth}\\
&& \deg(\omega_{\Itil\Jtil}) =-1 \comma \qquad \deg(\lamIJ)= +1 \comma \\
&& \deg(\mbox{rest})= 0  \period \label{deg1rest}
\end{eqnarray}
Under this grading, $\Qhat$ is decomposed into pieces 
 with degrees from $-1$ up to 6, with degree 4 missing. We  have 
\begin{eqnarray}
\Qhat &=& \deltatil + \Qtil + \dtil_1 + \dtil_2 + \dtil_3 + \dtil_5
+\dtil_6 \comma 
\end{eqnarray}
where $(\deltatil,\Qtil,\dtil_n)$, which carry degrees $(-1,0,n)$ 
respectively,  are given by (omitting the integral symbol $\int[dz]$
 for simplicity), 
\begin{eqnarray}
\deltatil &=& \half \lamIJ p_{\Itil \Jtil} \comma \\
\Qtil &=& -\lamp (p_- + 2i\delxp_I \thI) -\lamI (p_I + 2i \delxp_I \thp)
-ib_I\lamp \lamI -4c_\Itil \delxp_I \comma \\
\dtil_1 &=& \half \lam_{IJ} \left( 2i(\theta_\Itil \del x^-_\Jtil 
 -\theta_\Jtil \del x^-_\Itil ) + (\theta_\Itil \del \theta_\Jtil 
 -\theta_\Jtil \del \theta_\Itil)\theta_+ 
 -2\theta_\Itil \theta_\Jtil \del \theta_+ \right) \nn\\
&&\quad +4c_\Itil \lampinv \lam_{IJ} (\del x^-_\Jtil + i(\theta_+
\del \theta_\Jtil +\theta_\Jtil \del \theta_+)) 
 +4\lamp^{-2} c_\Itil c_\Jtil \lamIJ \del \thp \comma \\
\dtil_2 &=& -\lam_+ \theta_\Itil (\del \theta_{IJ} \theta_\Jtil
 -\theta_{IJ} \del \theta_\Jtil)+{i \over 8} \ep_{IJKLM}b_I\lam_{JK} \lam_{LM}
 \nn\\
&&\quad -\lam_\Itil (-2i\del x^-_\Jtil \theta_{IJ} + \del \theta_{IJ} \theta_\Jtil
\theta_+ -2\theta_{IJ} \del \theta_\Jtil \theta_+ 
+\theta_{IJ} \theta_\Jtil \del \theta_+ ) \nn\\
&&\quad +4ic_\Itil (\del\thIJ \thJ -\thIJ \del \thJ) -4\lampinv c_\Itil c_\Jtil
 \del \thIJ \comma \\
\dtil_3 &=& {i\over 2}  \ep_{IJKLM}\lam_{IJ} \theta_{KL} \del x^+_M \comma  \\
\dtil_5 &=& {1\over 4} \lamIJ   \ep_{IJKLM} \theta_{KL}(\del
 \theta_{MN} \theta_{\tilde{N}} 
-\theta_{MN}\del \theta_{\tilde{N}}) 
 +{1\over 4}\lamIJ  \ep_{JKLMN} \theta_\Itil 
 \theta_{KL}\del \theta_{MN}  \nn\\
&&\quad -ic_\Itil  \lampinv \lamIJ \ep_{JKLMN} \thKL \del \thMN  \comma\\
\dtil_6 &=& {1\over 4} \lamI \ep_{JKLMN}\thIJ \thKL \del \thMN \period
\end{eqnarray}
Although these expressions look complicated, what will become important 
 are the relatively simple relations among them which follow straightforwardly
 from the nilpotency of $\Qhat$: Decomposing $\Qhat^2=0$ 
with respect to the degree, we have 
\begin{eqnarray}
&& {\rm deg}= -2: \qquad \deltatil^2 =0 \comma \label{degmtwo}\\
&& {\rm deg}= -1: \qquad \deltatil \Qtil =0 \comma\label{degmone}\\
&& {\rm deg}= 0: \qquad \half \Qtil^2 +\deltatil \dtil_1 =0 \comma
\label{degzero}\\
&& {\rm deg}= 1: \qquad \Qtil \dtil_1 + \deltatil \dtil_2=0 \comma
\label{degone}\\
&& {\rm deg}= 2: \qquad \half \dtil_1^2 +  \Qtil \dtil_2 
+ \deltatil \dtil_3 =0 \comma\label{degtwo}\\
&& {\rm deg}= 3: \qquad \Qtil \dtil_3 + \deltatil \dtil_4 +\dtil_1 \dtil_2 
 =0 \comma\\
&& {\rm deg}= 4: \qquad \Qtil \dtil_4 + \deltatil \dtil_5 
+\half \dtil_2^2 + \dtil_3\dtil_1 =0 \comma\label{degfour}\\
&& {\rm deg}= 5: \qquad \Qtil \dtil_5 + \deltatil \dtil_6 +  \dtil_3 \dtil_2
 + \dtil_4 \dtil_1 =0 \comma\\
&& {\rm deg}= 6: \qquad \Qtil \dtil_6 + \half \dtil_3^2+\dtil_4 \dtil_2 
 +\dtil_5 \dtil_1 =0 \period
\end{eqnarray}

We can now clarify the sense in which the set of fields 
$(\lam_{IJ},\omega_{\Itil\Jtil}, \theta_{IJ}, p_{\Itil\Jtil})$ form a 
 quartet. From (\ref{degmtwo}) and (\ref{degmone}) we see 
 that $\deltatil$ is nilpotent and orthogonal to (\ie anticommutes with) $Q$. 
Further, it is trivial to check that $\deltatil \thIJ = \lamIJ$, 
 $\deltatil \lamIJ =0$,  $\deltatil \omega_{\Itil\Jtil}=p_{\Itil\Jtil}$ and
 $\deltatil p_{\Itil\Jtil}=0$. This clearly shows that the above set
 is a quartet with respect to a BRST-like opeartor $\deltatil$. 
Note also that, apart from $\deltatil$ itself, the members of the quartet 
 appear only in $d_n$'s with positive degrees. Thus, if we can remove 
 these $d_n$'s by a similarity transformation, we will be able to decouple 
 the quartet. This is exactly what we shall achieve below in several steps. 

First, consider the nilpotency relation (\ref{degzero})
 at degree 0. It is easy to check that actually each term of this equation 
 vanishes separately, \ie $\Qtil^2=0$ and $\deltatil \dtil_1=0$. The latter
 relation suggests that $\dtil_1$ may be written as $\deltatil \Rtil_2$ 
 for some degree 2 operator $\Rtil_2$. Since by inspection $\dtil_1$ 
is of the structure $\dtil_1=\half \lamIJ A_{IJ}$ such an operator
 is readily found:
\begin{eqnarray}
\Rtil_2 &=& \int[dz]\biggl( -4\,\imag \lampinv
    {{\theta }_+}
     \cIT{\tilde{I}}\thIJT{IJ}
     \partial \thIT{\tilde{J}}+
    {{\theta }_+}
     \thIT{\tilde{I}}\thIJT{IJ}
     \partial \thIT{\tilde{J}}+
    4\lamp^{-2}
    \cIT{\tilde{I}}
     \cIT{\tilde{J}}\thIJT{IJ}
     \partial {{\theta }_+}
    \brkeq +
    4\,\imag \lampinv
     \cIT{\tilde{I}}
      \thIT{\tilde{J}}\thIJT{IJ}
      \partial {{\theta }_+}-
    \thIT{\tilde{I}}
     \thIT{\tilde{J}}\thIJT{IJ}
     \partial {{\theta }_+}-4
    \lampinv\cIT{\tilde{I}}
     \thIJT{IJ}\partial \XmT{\tilde{J}} 
   -2\,\imag 
    \thIT{\tilde{I}}\thIJT{IJ}
    \partial \XmT{\tilde{J}}\biggr) \period
\end{eqnarray}

Next, we look at the nilpotency relation (\ref{degone}) at degree 1. 
Substituting $\dtil_1=\deltatil \Rtil_2$ and using $\Qtil \deltatil=0$ and 
 a Jacobi identity, 
 we have $0= \Qtil \dtil_1 
 + \deltatil \dtil_2 = \Qtil (\deltatil \Rtil_2) + \deltatil \dtil_2
= \deltatil (\dtil_2-\Qtil \Rtil_2)$. Just as before, this suggests 
 that there exists a degree 3 operator $\Rtil_3$ such that 
$\dtil_2-\Qtil \Rtil_2 =\deltatil \Rtil_3$ holds. After some computation, 
 we find that 
\begin{eqnarray}
\Rtil_3 \andeq -\frac{\imag}{8} \int[dz] 
   {{\epsilon }_{IJKLM}}
   \lamIJT{LM}\bIT{I}\thIJT{JK}  \comma 
\label{Rtil3}
\end{eqnarray}
satisfies the relation and hence $\dtil_2$ can be written as 
\begin{eqnarray}
 \dtil_2 &=& \Qtil \Rtil_2 + \deltatil \Rtil_3 \period \label{dtil2}
\end{eqnarray}

Let us go one more step to examine the relation (\ref{degtwo}) at degree 2. 
Since $\dtil_1^2=0$ holds by inspection, using (\ref{dtil2}), 
the nilpotency of $\Qtil$ and a Jacobi identity, we get
$0 =\half \dtil_1^2+ \Qtil \dtil_2+\deltatil \dtil_3 
= \Qtil (\Qtil \Rtil_2 + \deltatil 
 \Rtil_3) + \deltatil \dtil_3 = \deltatil (\dtil_3 -\Qtil \Rtil_3)$. 
By an explicit calculation, one finds that actually a stronger 
 relation $\dtil_3 =\Qtil \Rtil_3$ holds. 

At this point, one can already see a suggestive structure emerging. 
Using the expressions for $\dtil_1, \dtil_2$  and $\dtil_3$ obtained so far,
 $\Qhat$ can be rewritten as 
\begin{eqnarray}
\Qhat &=& \deltatil + \Qtil + \deltatil \Rtil_2 + \Qtil \Rtil_2 + \deltatil \Rtil_3 + \Qtil \Rtil_3 \cdots \nn\\
&=& \deltatil + \Qtil -\Rtil_2 (\deltatil + \Qtil) - \Rtil_3 (\deltatil 
+\Qtil) + \cdots  \nn\\
&=& (1-\Rtil_2-\Rtil_3) (\deltatil + \Qtil) + \cdots \period
\end{eqnarray}
This is recognized as the beginning of a similarity transformation of
 the form $\Qhat = e^{-(\Rtil_2+\Rtil_3+ \cdots)} (\deltatil + \Qtil) 
 e^{\Rtil_2 + \Rtil_3+ \cdots}$. 

In fact, similar but more involved 
 analysis of the nilpotency  relations at higher degrees confirms 
that this pattern continues to hold and terminates after  finite 
 steps, although multiple actions of $\Rtil_i$'s occur 
in non-trivial ways starting at degree 5. 
Omitting the details, the final answer is given by 
\begin{eqnarray}
\Qhat &=& e^{-\Rtil} (\deltatil +\Qtil )e^{\Rtil} \comma \\
\Rtil &=& \Rtil_2 + \Rtil_3 + \Rtil_5 +\Rtil_6 + \Rtil_8 + \Rtil_9 \comma 
\end{eqnarray}
where, suppressing $\int[dz]$, 
\begin{eqnarray}
 \Rtil_2\andeq -4\,\imag \lampinv
    {{\theta }_+}
     \cIT{\tilde{I}}\thIJT{IJ}
     \partial \thIT{\tilde{J}}+
    {{\theta }_+}
     \thIT{\tilde{I}}\thIJT{IJ}
     \partial \thIT{\tilde{J}}+
    4\lamp^{-2}
    \cIT{\tilde{I}}
     \cIT{\tilde{J}}\thIJT{IJ}
     \partial {{\theta }_+}
    \brkeq +
    4\,\imag \lampinv
     \cIT{\tilde{I}}
      \thIT{\tilde{J}}\thIJT{IJ}
      \partial {{\theta }_+}-
    \thIT{\tilde{I}}
     \thIT{\tilde{J}}\thIJT{IJ}
     \partial {{\theta }_+}-4
    \lampinv\cIT{\tilde{I}}
     \thIJT{IJ}\partial \XmT{\tilde{J}} \brkeq 
   -2\,\imag 
    \thIT{\tilde{I}}\thIJT{IJ}
    \partial \XmT{\tilde{J}}\comma \\
\Rtil_3 \andeq -\frac{\imag}{8}
   {{\epsilon }_{IJKLM}}
   \lamIJT{LM}\bIT{I}\thIJT{JK}\comma\\
\Rtil_5 \andeq \frac{1}{12}\lampinv
    {{\epsilon }_{IJKLM}}
    \lamIJT{JK}
    {{\theta }_+}\thIJT{IN}
     \thIJT{LM}
     \partial \thIT{\tilde{N}}-
     \frac{\imag}{6}\lamp^{-2}
    {{\epsilon }_{IJKLM}}
    \lamIJT{JK}
    \cIT{\tilde{N}}\thIJT{IN}
     \thIJT{LM}
     \partial {{\theta }_+}
    \brkeq +\frac{1}{12}\lampinv
    {{\epsilon }_{IJKLM}}
    \lamIJT{JK}
    \thIT{\tilde{N}}\thIJT{IN}
     \thIJT{LM}
     \partial {{\theta }_+}-
     \frac{\imag}{12}\lampinv
    {{\epsilon }_{IJKLM}}
    \lamIJT{JK}
    \thIJT{IN}\thIJT{LM}
    \partial \XmT{\tilde{N}}\comma
    \\
\Rtil_6 \andeq \frac{\imag }{3}\lampinv
    {{\epsilon }_{IJKLM}}
    \cIT{\tilde{N}}\thIJT{IJ}
     \thIJT{MN}
     \partial \thIJT{KL}-
    \frac{1}{12}
    {{\epsilon }_{IJKLM}}
    \thIT{\tilde{N}}\thIJT{IJ}
     \thIJT{MN}
     \partial \thIJT{KL}\comma\\
\Rtil_8 \andeq \frac{1}{480}\lamp^{-2}
   {{\epsilon }_{IJKLM}}
   {{\epsilon }_{NPQRS}}
   \lamIJT{LM}\lamIJT{NS}
   \thIJT{IP}\thIJT{JK}\thIJT{QR}
    \partial {{\theta }_+}\comma\\
\Rtil_9 \andeq \frac{1}{240}\lampinv
   {{\epsilon }_{IJKLM}}
   {{\epsilon }_{NPQRS}}
   \lamIJT{JM}
   \thIJT{IN}\thIJT{KL}\thIJT{PQ}
    \partial \thIJT{RS} \period
\end{eqnarray}
(In deriving this result, one needs some non-trivial identities 
among several quantities of the type appearing in $\Rtil_8$. )

Thus, we have succeeded in reducing our original $\Qhat$ to the
 sum of mutually orthogonal nilpotent  operators $\deltatil$ and $\Qtil$, 
 where the former acts only on the space of the quartet and 
 the latter on the rest of the fields. 
 This shows that
 the cohomology of $\deltatil$, which is trivial as already argued,
 is decoupled and cohomologically $\Qhat$ is equivalent to $\Qtil$.

 Evidently, our similarity transformations are well-defined
for $\lambda_+\ne 0$, just as in Berkovits' formalism. However,
 the apparent singularity at $\lambda_+=0$ is just a coordinate singularity
 in the $\lambda$-space and does not affect the cohomology, which
 is known to be Lorentz invariant \cite{Berk01051}.

\subsection{Decoupling of the second quartet}
Having reduced $\Qhat$ down to $\Qtil$, we now decouple the second quartet 
$(b_I,c_\Itil ,\omega_I, \lamI)$ from the cohomology of $\Qtil$. 
This can be achieved quite analogously as above. Let us assign 
 new non-vanishing degrees to the members of the quartet as follows:
\begin{eqnarray}
&& \deg(b_I)= -2\comma  \qquad \deg(c_\Itil)= +2 \comma \label{deg2bc}\\
&& \deg(\omega_I) =-1 \comma \qquad \deg(\lamI)= +1 \comma \\
&& \deg(\mbox{rest})= 0  \period \label{deg2rest}
\end{eqnarray}
Then, $\Qtil$ is decomposed as 
\begin{eqnarray}
\Qtil &=& \deltabar + \Qbar + \dbar_1 + \dbar_2 \comma 
\end{eqnarray}
where $(\deltabar, \Qbar, \dbar_1, \dbar_2)$, carrying the degrees 
 $(-1,0,1,2)$,  are given by 
\begin{eqnarray}
\deltabar &=& -i \int[dz] \lamp b_I \lamI \comma \\
\Qbar &=& -\int[dz] \lamp \dhat_- \comma \\
\dbar_1 &=& -\int[dz] \lamI \dhat_I \comma \\
\dbar_2 &=& -4\int[dz] c_\Itil \del \xp_I \period
\end{eqnarray}
In the above,  $\dhat_-$ and $\dhat_I$ are defined as
\begin{eqnarray}
\dhat_- &=& p_- + 2i \del \xp_I \thI \comma \\
\dhat_I &=& p_I +2i \delxp_I \thp \period 
\end{eqnarray}
Again from the nilpotency $\Qtil^2=0$, the relations formally 
 similar to (\ref{degmtwo}) $\sim$ (\ref{degfour}) follow. 
 Actually these relations 
reduce in this case to nilpotency of each operator and to 
 simple anticommutation relations among them,
 except for one non-trivial relation at degree 1 given by
\begin{eqnarray}
\Qbar \dbar_1 + \deltabar \dbar_2 = 0 \period \label{Qbardbar1}
\end{eqnarray}
Now the relation $\deltabar \dbar_1=0$ suggests that $\dbar_1$ can be 
 expressed as $\dbar_1=\deltabar S_2$, 
 with some operator $S_2$ of degree 2. It is easily found to be
given by 
\begin{eqnarray}
S_2 &=& -i\int[dz] \lampinv c_\Itil \dhat_I \period
\end{eqnarray}
Putting this result into (\ref{Qbardbar1}) we get 
 $\deltabar (\dbar_2 +S_2\Qbar ) =0$. In fact by simple calculations
one can check the following properties of $S_2$:
\begin{eqnarray}
S_2 \Qbar &=& -\dbar_2\comma \qquad S_2 \dbar_2 =0 \comma 
\qquad S_2 \dbar_1 = 0\period
\end{eqnarray}
These relations are sufficient to verify the validity of 
 the similarity transformation 
\begin{eqnarray}
\Qtil &=& e^{-S_2} (\deltabar + \Qbar) e^{S_2} \period 
\end{eqnarray}
Therefore, just as in the previous subsection, the set of fields 
$(b_I,c_\Itil ,\omega_I, \lamI)$  form a quartet with respect to 
 the nilpotent operator $\deltabar$ and, as $\deltabar$ and $\Qbar$ 
are mutually orthogonal, they are decoupled from the physical sector 
 governed by the cohomology of $\Qbar$. 
 It should also be noted that 
 under this similarity transformation,  $\deltatil$, which played a key role 
 in the previous subsection, is unaffected. 

What is rather remarkable is that the information of the 
non-trivial cohomology in EPS and hence in 
 PS formalism is contained in a drastically simplified nilpotent operator 
\begin{eqnarray}
\Qbar &=& -\int[dz] \lamp \dhat_- = -\int[dz] 
\left(\lamp p_- +2i \lamp \delxp_I \thI \right)\period
\end{eqnarray}
In the next section, we shall show that this operator is connected 
 to the BRST charge  of the conventional RNS formalism
in the large Hilbert space 
by another similarity transformation. 
\subsection{Reduction of the $B$-ghost field}
Having decoupled the two quartets and transformed $\Qhat$ to a simple operator
 $\Qbar$, it is of interest to see how the  $B$-ghost given in 
(\ref{Bghost}) gets transformed by the similarity transformation. 
This analysis will turn out to shed light on the reason 
why it is difficult to construct its couterpart in the PS formalism
 formulated in  smaller field space. 

According to the grading 
 introduced in (\ref{deg1pth}) $\sim$ (\ref{deg1rest}), $B$ given in 
 (\ref{Bghost}) is decomposed into the following three pieces with 
 designated degrees:
\begin{eqnarray}
B &=& B_0 + B_1 + B_4 \comma \\
B_0 &=& \half b_I ( \delxm_\Itil + i(\thp \del\thI + \thI \del \thp))
 + \omega_- \del\thp + \omega_I \del\thI \comma \\
B_1 &=& -\half \omega_{\Itil\Jtil} \del\thIJ \comma \\
B_4 &=& -{1\over 8} \ep_{IJKLM} b_I \theta_{\Jtil\Ktil}
\del\theta_{\tilde{L}\tilde{M}} \period
\end{eqnarray}
It is not difficult to show that under the first similarity transformation
 $e^{\Rtil} (\ast) e^{-\Rtil}$, $B$ is turned into
\begin{eqnarray}
e^{\Rtil} B e^{-\Rtil} &=& \Btil \equiv B_0 + B_1 \period
\end{eqnarray}
In other words, the effect of the similarity transformation is simply 
 to remove the piece $B_4$. 

It should now be noted that by any similarity transformation of 
 the form $e^W (\ast) e^{-W}\comma W =\int[dz] j(z)$, with $j(z)$ 
 a primary field of dimension 1, the energy momentum tensor $T(w)$
 is unchanged. This is because $W T(w) = \int[dz] j(w)/(z-w)^2 =0$. 
Due to this property, we must have  $(\deltatil+\Qtil) (\Btil_0
 + \Btil_1) = T_{EPS}$. Indeed, we find the nice relations
\begin{eqnarray}
\deltatil B_1 &=& T_{(p,\theta, \omega, \lam)} \comma \\
\Qtil B_0 &=& T_{EPS} -T_{(p,\theta, \omega, \lam)} \comma \\
\deltatil B_0 &=& \Qtil B_1 =0 \comma
\end{eqnarray}
where $T_{(p,\theta, \omega, \lam)}$ is the energy-momentum tensor 
 for the first quartet $(\lam_{IJ},\omega_{\Itil\Jtil}, \theta_{IJ}, p_{\Itil\Jtil})$. This shows that $B_0$ acts as the proper 
 $B$-ghost in the space without the first quartet, where $\Qtil$ serves as
 the BRST charge. Next we consider the effect of the second similarity 
 transformation $e^{S_2} (\ast) e^{-S_2}$. Under the second grading 
 (\ref{deg2bc}) $\sim$ (\ref{deg2rest}),  $B_0$ above is split as
\begin{eqnarray}
B_0 &=& \Btil_{-2} + \Btil_{-1} + \Btil_0 \comma 
\end{eqnarray}
where 
\begin{eqnarray}
\Btil_{-2} &=& \half b_I ( \delxm_\Itil + i(\thp \del\thI + \thI \del \thp))
\comma \\
\Btil_{-1} &=& \omega_I \del\thI \comma \\
\Btil_0 &=& \omega_- \del \thp 
\period
\end{eqnarray}
A straightforward computation produces the structure
\begin{eqnarray}
e^{S_2} B_0 e^{-S_2} &=& \Bbar_{-2} + \Bbar_{-1} + \Bbar_0 + \Bbar_1 \comma 
\end{eqnarray} 
where $\Bbar_{-2}= \Btil_{-2}$,  $\Bbar_{-1} = \Btil_{-1}$, 
 $\Bbar_0 = \Btil_0 + S_2 \Btil_{-2}$,   and 
 $\Bbar_1 = S_2 \Btil_{-1}$. Since $\Qtil$ is transformed into $\deltabar 
+\Qbar$, we must have $(\deltabar + \Qbar) (\Bbar_{-2} + \Bbar_{-1} + \Bbar_0 + \Bbar_1) = T_{EPS} -T_{(p,\theta, \omega, \lam)}$. In fact, the 
 non-vanishing contributions on the LHS are found to be 
\begin{eqnarray}
\deltabar \Bbar_1 &=& T_{(b,c,\omega,\lam)} + \left( b_Ic_\Itil 
 \del s + {5\over 2} ((\del s)^2 -\del^2 s)\right) \comma \\
\Qbar \Bbar_0 &=& T_{EPS} - T_{(p,\theta, \omega, \lam)} -T_{(b,c,\omega,\lam)}
-\left( b_Ic_\Itil 
 \del s + {5\over 2} ((\del s)^2 -\del^2 s)\right) \comma 
\end{eqnarray}
where $T_{(b,c,\omega,\lam)}$ is the energy-momentum tensor for the 
 second quartet $(b_I,c_\Itil ,\omega_I, \lamI)$ and 
 the bosonic field $s$ is defined by $\lamp=e^s$. Therefore, 
 although the sum correctly reproduces $T_{EPS} -T_{(p,\theta, \omega, \lam)}$,
$\Bbar_0$ cannot be regarded as the $B$-ghost for 
 the PS formalism. This can be taken as a strong indication 
that in the constrained  field space an appropriate $B$-ghost field 
cannot be constructed. 
\section{Similarity Transformation in RNS }
\subsection{Preliminary}
Having decoupled the extra quartets in the EPS,  the 
 degrees of freedom now match precisely to the ones in the RNS formulated 
 in the large Hilbert space $\calH_L$. As was shown in \cite{Berk0104}, 
in $\calH_L$ the physical spectrum is characterized by the cohomology 
 of the extended BRST operator\footnote{This characterization is valid 
 provided that a finite range of ``pictures'' are used \cite{Berk0104}.
This point will be elaborated in Sec.~5.2, 
 where we discuss the issue of the correct cohomology. } 
\begin{eqnarray}
\QpR &=& \eta_0 + \QR \comma 
\end{eqnarray}
where $\eta_0$ is the zero mode of $\eta$ and $\QR$ is the usual 
 BRST operator 
\begin{eqnarray}
Q_{RNS} &=&\int[dz]\left[ cT_M -\half e^\phi \eta G_M +bc\del c 
-c \left( \half (\del\phi)^2 +\del^2\phi +\eta \del \xi\right)
-{1\over 4}e^{2\phi} b\eta \del\eta 
\right] \period \nn\\
\end{eqnarray}
Here $T_M$ and $G_M$ are, respectively, 
 the energy-momentum tensor and the superconformal
 generator for the matter sector given by 
\begin{eqnarray}
T_M &=& -\half \del x^\mu \del x_\mu -\half \psi^\mu \del \psi_\mu \comma\\
G_M &=& i \del x^\mu \psi_\mu \period
\end{eqnarray}
Our goal in this section is to try to find a similarity transformation 
 which transforms $\QpR$ into the simple nilpotent operator $\Qbar$,
obtained in the previous section, under appropriate identification of 
 fields of EPS and RNS. 

Before we begin the consruction, we should mention that in the past 
an example of a drastic simplification of $\QR$ by a similarity 
transformation has been noted \cite{Acosta:1999hi}. 
Namely, it was found that 
\begin{eqnarray}
e^W \QR e^{-W} &=& \int[dz] \left( -{1\over 4} e^{2\phi} b\eta \del\eta 
\right) \comma 
\end{eqnarray}
where 
\begin{eqnarray}
W &=& W_1 + W_2 \comma \\
W_1 &=& 2i \int[dz] e^{-\phi} c \xi \psi_\mu \del x^\mu \comma \label{Wone}\\
W_2 &=& -2\int[dz] \del \phi  e^{-2\phi} c\del c \xi \del \xi \period
\label{Wtwo}
\end{eqnarray}
This remarkable representation found some applications in the context of 
superstring field theory \cite{SSFT}. At the same time, however, 
under this transformation 
$\eta_0$ turns into a complicated expression\footnote{To our knowledge, 
this expression has not been recorded in the literature.}
\begin{eqnarray}
&& \eta_0 + \int[dz] \Bigl[
 - 2i e^{-\phi}c\psi_{\mu}\delx^{\mu}
 + 2e^{-2\phi}c\del c\xi
     \bigl(\delx_{\mu}\delx^{\mu} + \psi_{\mu}\del\psi^{\mu}\bigr) \nonumber\\
&&\quad +e^{-2\phi}\Bigl(
 10c\del c\xi(\del\phi)^2
 -8c\del c\del\xi \del\phi
 -{10\over3}c\del^3c\xi\Bigr) \Bigr]
\end{eqnarray}
 so that $\QpR$ as a whole is not simplified. Thus, we must seek 
 a different transformation. 
\subsection{First step}
Let us now describe our construction. It will be done in two steps, again 
by  introducing judicious gradings and making use of the relations that follow 
 from the nilpotency of the BRST charge. 

As the first step,
 we adopt the bosonized representation of the $\be$-$\ga$ ghosts 
 and assign to the fields the following degrees:
\begin{eqnarray}
\deg( \eta,\xi) &=& (1,-1) \comma \quad 
\deg( c,b) = (5,-5) \comma \\
\deg(\psip_I,\psim_I) &=& (2,-2)\comma\quad  \deg (e^{n\phi}) =n\comma 
\qquad \deg(\mbox{rest}) =0 \period
\end{eqnarray}
Then, $\QpR$ decomposes into five terms as 
\begin{eqnarray}
\QpR &=& \delta  + \Qp +\eta_0 + \Qm + d \comma \\
\delta &\equiv & -{1\over 4} b\eta \del \eta e^{2\phi}\comma  \\
\Qp &\equiv & -\half e^\phi \eta G_M^+ \comma \\
\Qm &\equiv & -\half e^\phi \eta G_M^- \comma \\
d &\equiv & c\left( T_M -\half (\del\phi)^2 -\del^2\phi -\eta \del \xi
\right) + bc\del c \comma
\end{eqnarray}
where $G_M^\pm$ are defined as 
\begin{eqnarray}
G_M &=& G_M^+ +G_M^- \comma \\
G_M^+ &=& 2i \psi^-_\Itil \delxp_I\comma \qquad G_M^-
 = 2i \psi^+_I \delxm_\Itil 
\period 
\end{eqnarray}
The operators $(\delta, \Qp,\eta_0, \Qm, d)$ carry degrees 
$(-1, 0,1,4,5)$ respectively. From the nilpotency of $\QpR$ and $\QR$ 
 we easily find that except for one non-trivial relation 
\begin{eqnarray}
\Qp \Qm + \delta d =0 \comma\label{QpQm}
\end{eqnarray}
all the five operators are nilpotent and anticommute with each other. 
In particular, the relation $\delta \Qp =0$ suggests that $\Qp$ can 
 be written as 
\begin{eqnarray}
\Qp &=& \delta \T \comma \label{QpT}
\end{eqnarray}
with some operator $\T$ of degree 1. Such an operator is easily found 
 to be given by
\begin{eqnarray}
\T &=& 2\int[dz]c \xi e^{-\phi} G^+_M 
=4 i\int[dz] c\xi e^{-\phi} \psim_\Itil \delxp_I \period
\end{eqnarray}
It is intriguing to note that this operator is precisely ``half'' of 
$W_1$ given in (\ref{Wone}). 
For us the importance of this operator is that when acting on $\eta_0$ 
 it produces 
\begin{eqnarray}
\T \eta_0 &=& Q_0 \equiv 4i\int[dz] ce^{-\phi} \psim_\Itil \delxp_I \comma 
\end{eqnarray}
which will eventually be identified with the second 
piece $-2i\lamp \thI \delxp_I$ of $\Qbar$ in EPS formalism. Moreover, 
 since $\T\T \eta_0 =0$, the following similarity transformation holds:
\begin{eqnarray}
e^{\T} \eta_0 e^{-\T} = \eta_0 + Q_0 \period \label{simTeta}
\end{eqnarray}
In fact, as we shall later identify $\eta_0$ with the first term $-\lamp p_-$
 of $\Qbar$, the RHS of (\ref{simTeta}) will become nothing but 
$\Qbar$ itself. At this point  of the analysis, however, it 
is not yet of great significance 
 since this is only a small part of the similarity transformation and we still 
 have many terms left to be transformed. 

Let us study the consequence of the relation (\ref{QpQm}) using the 
representation (\ref{QpT}).  Since $\delta \Qm=0$,  it can be rewritten as
\begin{eqnarray}
0&=& (\delta \T)\Qm + \delta d = \delta (\T \Qm + d) \period
\end{eqnarray}
This suggests that $\T \Qm +d$ can be written as 
\begin{eqnarray}
\T \Qm +d &=& \delta X \comma \label{deltaX}
\end{eqnarray}
for some $X$ of degree 6. By an explicit calculation of the LHS, it is 
 not difficult to show that $X$ is given by 
\begin{eqnarray}
X &=& \int[dz] (4\psim_\Itil \psip_I -2\del \phi)
 e^{-2\phi} c\del c \xi \del \xi 
\period 
\end{eqnarray}
Again, curiously the second half of this operator is identical to $W_2$
 shown in (\ref{Wtwo}). 

We are now in a position to look at how the rest of the terms in $\QpR$ 
are transformed under the similarity transformation $e^{T} (*) e^{-T}$. 
The commutation relations required for this purpose are easily computed 
as 
\begin{eqnarray}
T\Qp &=& 0 \comma \quad TT\Qm = -2Td\comma  \quad TTd =0 \comma \\
TX &=& \Qm X = d X= 0\comma \quad \Qp X = Td \period \label{relX}
\end{eqnarray}
They are enough to lead to 
\begin{eqnarray}
e^T \Qp e^{-T} &=& \Qp \comma \\
e^T \Qm e^{-T} &=& \Qm -d + \delta X -Td \comma \\
e^T \delta e^{-T} &=& \delta -\Qp \comma \\
e^T d e^{-T} &=& d+Td \comma 
\end{eqnarray}
and, together with the transformation of $\eta_0$ already discussed 
in (\ref{simTeta}), we obtain 
\begin{eqnarray}
e^T \QpR e^{-T} &=& \eta_0 + Q_0 + \delta + \Qm + \delta X \period 
\end{eqnarray}
Although we do not display it here, the explicit form of 
the last term, $\delta X$, is rather complicated and it is desirable to 
 remove it before moving on to the next step. This can be achieved 
 by the similarity transformation of the form $e^X (*) e^{-X}$, although 
 it produces an additional term  
\begin{eqnarray}
\dch_1  &\equiv & e^X \eta_0 e^{-X} =-\int[dz]
(4\psim_\Itil \psip_I -2\del \phi)
 e^{-2\phi} c\del c  \del \xi \period
\end{eqnarray}
In this way, by using the relations given in (\ref{relX}) and (\ref{deltaX}),
 one arrives at 
\begin{eqnarray}
e^X e^T \QpR e^{-T} e^{-X} &=& \QchR\equiv 
\eta_0 + Q_0 + \delta + \Qm + \dch_1 \period
\end{eqnarray}
%
\subsection{Second step}
Now we proceed to the second step and show that $\QchR$ can be brought 
 precisely to the form $\eta_0 + Q_0$ by a further similarity transformation. 

To this end, we shall introduce yet another grading scheme and assign 
to the fields the following degrees:
\begin{eqnarray}
\deg(\eta,\xi) &=& (-1, 1) \comma \qquad  \deg(c,b) = (4,-4) \comma \\
\deg \left(e^{\phi} \right) &=& 4 \comma \qquad 
\deg (\mbox{rest}) =0 \period
\end{eqnarray}
Then $\QchR$ is decomposed as 
\begin{eqnarray}
\QchR &=& \deltach + \Qch_0 + \dch_1 +\dch_2+\dch_3 \comma 
\end{eqnarray}
where $(\deltach, \Qch_0, \dch_1,\dch_2,\dch_3)$ which carry 
 degrees $(-1,0,1,2,3)$ respectively are given by 
\begin{eqnarray}
\deltach &=& \eta_0\comma  \\
\Qch_0 &=& 2\int[dz]ce^{-\phi} G^+_M = Q_0\comma  \\
\dch_1 &=&  - \int[dz](4\psi^-_I \psi^+_I -2\del \phi) 
e^{-2\phi}  c \del c  \del \xi \comma \\
\dch_2 &=& -{1\over 4}\int[dz] b\eta\del\eta e^{2\phi} =\delta  \comma \\
\dch_3 &=&  -\half\int[dz] e^\phi \eta G^-_M = \Qm\period
\end{eqnarray}
Obviously the new grading merely reorders the previous operators in 
 a convenient way. 
These operators are all nilpotent and anticommute with each other,
 except for one non-trivial relation 
\begin{eqnarray}
\Qch_0 \dch_3 + \dch_1 \dch_2 &=& 0 \comma \label{Qch0dch}
\end{eqnarray}
which follows from $\QchR^2=0$. 

As we wish to remove $\dch_2$, let us focus on the relation $\Qch_0 \dch_2=0$.
By the reasoning repeatedly used, we can find an operator $Y_2$ of degree 2
such that 
\begin{eqnarray}
\dch_2 &=& \Qch_0 Y_2 \period \label{dch2}
\end{eqnarray}
The explicit form of $Y_2$ is 
\begin{eqnarray}
Y_2 &=&-{i \over 20} \int[dz]e^{3\phi}  b\del b \eta \del \eta \psi^+_I x^-_\Itil
\period
\end{eqnarray}
Subsitituting (\ref{dch2}) into (\ref{Qch0dch}) and rearranging, we immediately
 find $\Qch_0 (\dch_3+Y_2 \dch_1) =0$. This implies a relation of the 
 form 
\begin{eqnarray}
\dch_3+Y_2 \dch_1 &=& \Qch_0 Y_3 
\end{eqnarray}
for some operator $Y_3$ of degree 3 and indeed it is given by 
\begin{eqnarray}
Y_3 &=& -{1 \over 5}\int[dz] e^{2\phi} b\eta (\psip_Ix^-_\Itil) (\psip_J \delxm_\Jtil) 
\period
\end{eqnarray}

With this preparation,  we now examine a similarity transformation 
of the form $e^Y (*) e^{-Y}$
 with $Y=Y_2 +Y_3$. It is straightforward to check the relations
\begin{eqnarray}
Y (\dch_2+\dch_3) &=& 0 \comma \qquad Y (\Qch_0 + \dch_1) = -(\dch_2+\dch_3) 
\comma \qquad Y \deltach =0\comma 
\end{eqnarray}
and from this  we easily get 
\begin{eqnarray}
e^Y \QchR e^{-Y} &=& \deltach + \Qch_0 + \dch_1 
= \eta_0 + Q_0 + \dch_1 \period
\end{eqnarray}

Finally, let us remove $\dch_1$. This is done simply by the inverse 
 similarity transformation using the operator $X$, since 
$e^{-X} (\eta_0 + \dch_1) e^{X} = \eta_0$ and $X Q_0 =0$. 

Summarizing, after a rather long but systematic procedure, we have 
established a desired formula 
\begin{eqnarray}
e^{-X} e^{Y} e^{X} e^{T} \QpR e^{-T} e^{-X} e^{-Y} e^{X} 
 = \eta_0 + Q_0 \period
\end{eqnarray}
In the next section, we shall show that the RHS
 precisely matches 
 the operator $\Qbar$ on the EPS side, as promised. 
\section{Mapping between EPS and RNS}
\subsection{Identification of fields and BRST operators}
To identify the simplified BRST operators on the EPS and the RNS sides, 
it is necessary to map the basic fields of these formalisms. Fortunately, such 
 a mapping was already proposed by Berkovits in \cite{Berk0104} and 
 essentially we only need to make use of this scheme with minor modifications. 

Before we give the explicit identification rules, we wish to make 
 a remark. Although a similarity transformation induces redefinition 
 of fields,  the identification rules are form-invariant: Both sides 
 of the relations are transformed in the same way so that the OPE's are 
 retained. Thus, we shall find that the conversion rules in \cite{Berk0104}, 
which were applied on RNS side {\it before} various manipulations, 
remain correct in our case where two theories are connected ``in the middle''
 {\it after}  application of similarity transformations  on both sides.

The mapping is best described using the ``bosonized'' form of various 
 quantities. Besides the ones  already described, 
 we introduce, as in \cite{Berk0104}, a pair of conjugate 
 bosons $(s,t)$ with the OPE
\begin{eqnarray}
s(z) t(w) &=& \ln (z-w)\comma \qquad s(z) s(w) = t(z)t(w) =0 \comma
\end{eqnarray}
and the energy-momentum tensor
\begin{eqnarray}
T_{st} &=& \del s \del t + \del^2 s \period
\end{eqnarray}
Then,  $\lamp$ and its conjugate $\omega_-$ can be expressed as 
\begin{eqnarray}
\lamp &=& e^s\comma \qquad \omega_- = \left( \half \del s + \del t\right)
e^{-s} \period
\end{eqnarray}
It is easy to check that the OPE as well as the energy-momentum tensor
 for $(\lamp, \omega_-)$  are correctly reproduced. 

 The basic mapping can then be described as follows\footnote{Originally, 
 $\thp,p_-,\thI$ and $p_I$ are introduced as independent part of components 
 of the spin fields $\Sigma^\al$ and $\Sigma_\al$ in appropriate pictures
  \cite{Berk0104}. 
We find it more convenient to display the quantities multiplied by 
 the factors $e^{\pm s}$ in order to avoid the non-trivial 
 Jordan-Wigner factors \cite{Kosteleckyetal}
 associated with these spin fields.}:
\begin{eqnarray}
-e^s p_- &=& \eta \comma \qquad e^{-s} \thp = \xi \comma \label{map1}
\\
e^t \thp &=&  c 
 \comma \qquad e^{-t} p_- = - b 
\comma  \label{map2} \\
e^{-s} p_I &=&  -be^\phi  \psip_I \comma \qquad 
e^s \thI = -2 c e^{-\phi}  \psim_\Itil \period \label{map3}
\end{eqnarray}
It is easy to check that they reproduce the correct OPE's on both sides. 

A further non-trivial check of the rules above is provided by the correct 
 conversion 
 of the energy-momentum tensors. To demonstrate it, it is convenient 
 to bosonize $\xi, \eta, c, b$ and $\psi^\pm_I$ as in (\ref{bosxieta}), 
 (\ref{boscb}) and (\ref{bospsiI}). 
Then, one can express $(s,t)$ bosons in terms of RNS bosons as \cite{Berk0104}
\begin{eqnarray}
s &=& \sigma -{3\over 2} \phi + \half H \comma \qquad
t = -\chi + {3\over 2}\phi -\half H\period 
\end{eqnarray}
Now let us sketch how one can convert the part of $T_{EPS}$ shown in 
 (\ref{TEPS}), with the two quartets dropped, into $T_{RNS}$ by using 
 the correspondence rules above. 
Since $s$ does not have singular OPE's with $\eta$ nor $\xi$, one can 
easily express $p_-$ and $\thp$ in terms of RNS variables using 
 (\ref{map1}), and $p_-\del \thp$ can be readily computed.
 On the other hand, similar manipulations cannot be 
 applied to (\ref{map3}) as $s$ does have singular OPE's with the RHS. 
However, this can be gotten around by 
combining the finite parts of $(e^{-s} p_I)(z) \del (e^s \thI)(w)$ and 
 $(e^{-s} p_I)(z)  (e^s \thI \del s)(w)$, which can be computed
 easily. In this way one can express $p_I \del \thI$ in terms of 
the RNS bosons. Adding in the RNS expression for $T_{st}$, many cancellations
 take place and we indeed reproduce $T_{RNS}$.

Having established the identification rules, 
 we can easily compare the BRST operators $\Qbar$ for EPS 
and $Q_0$ for RNS, obtained in the previous section. Recall the 
 form of $\Qbar$:
\begin{eqnarray}
\Qbar &=&\int[dz]\left( -\lamp p_- -2i \lamp\thI \delxp_I \right) =
\int[dz]\left( -e^s p_- 
 -2i  e^s \thI \delxp_I\right) \period
\end{eqnarray}
Applying the map (\ref{map3}), we see that this is nothing 
 but $Q_0 = \eta_0 + 4i\int[dz] ce^{-\phi} \psim_\Itil \delxp_I$ 
 and hence the BRST operator for the EPS is directly connected 
 by a series of similarity transformations to the one for the RNS,
 modulo two quartets which cohomologically decouple. 

We wish to emphasize that, 
 in contrast to the corresponding procedure developed by Berkovits for 
the PS formalism, our transformations do not 
 involve any singular operations or functions. Evidently, this must be 
 due to the use of extended field space, without 
 the PS constraints, in the case of our formalism. As explained in 
 Sec.~3 and 4, construction of our similarity transformations appears
 very natural, following essentially from the nilpotency structure of the 
BRST charges. 

\subsection{Proper Hilbert space and cohomology}
Although we have succeeded in connecting the EPS and the RNS formalisms 
 by means of a similarity transformation, there still remains an important 
 question of the proper Hilbert space in which to consider the cohomology. 

The generic problem is as follows: Suppose that there exists 
 a local fermionic operator $\Xi(z)$ which is ``inverse''
to the BRST operator $Q$ 
 in the sense $Q\Xi(z) =1$. Then any BRST-closed operator $V(z)$ 
 can always be written as a BRST-exact form $V(z) = Q(\Xi(z)V(z))$, 
since $Q(\Xi(z)V(z))
 = (Q\Xi(z))V(z) -\Xi(z) (QV(z)) = V(z)$. Hence in such a situation 
 the cohomology of $Q$ becomes trivial. As was noted by Berkovits 
\cite{Berk0006,Berk0104}, such an 
 operator indeed exists in PS formalism 
and is given (up to an irrelevant overall scale 
 and a $Q$-exact term) by $\Xi = \lamp^{-1}\thp$.  This operator continues
 to be the inverse to our BRST operator $\Qhat$ in EPS as well. 
The most natural way to disallow such 
 an operator is, as was postulated by Berkovits \cite{Berk0104}, 
to limit the Hilbert space to the so-called
 ASPC (almost super-Poincar\'e covariant) subspace. Namely, one allows 
only those operators which transform covariantly under the spacetime 
 SUSY and $U(5)$ subgroup of the super-Poincar\'e group. Then since 
 the vertex operators $V$  constructed in (E)PS are known to have 
ASPC representatives, the products 
 $\Xi V$ are not ASPC (due to SUSY-non-invariance of $\theta_+$) and
 hence are excluded. Evidently the notion of ASPC is still robust 
 upon similarity transformations on the EPS side, although 
 the form of the supercharges get modified. It will be useful to 
 note that the operator $\Xi=\lampinv\thp$,
 on the other hand,  can be checked to be
 form-invariant under these transformations.

Now since the purpose of our work is to relate EPS to RNS, we must also
understand how this restriction of the Hilbert space is justified 
 from the point of view of RNS formalism. Let us recall that (E)PS 
 formalism is connected to the RNS formalism in the 
 ``large'' Hilbert space ${\cal H } _l $ with $\xi_0$ mode, where 
 the extended BRST charge is given by $Q'_{RNS} = \eta_0 + Q_{RNS}$. 
It was demonstrated in \cite{Berk0104} that 
 the cohomology of $Q'_{RNS}$ is equivalent to the conventional cohomology of 
$Q_{RNS}$  in the ``small'' Hilbert space without $\xi_0$, 
provided that ${\cal H}_l$
 is restricted to the space of operators with finite range of 
 pictures, to avoid triviality of the cohomology. This implies 
 that if the operator inverse to $Q'_{RNS}$ exists,
 it must carry infinite range of pictures. Such an operator has not been 
 identified previously,
 although the inverses to $\eta_0$ and $Q_{RNS}$ separately
 are well-known. 

Let us now analyze the nature of the operator $\Xi$ in RNS. According 
 to the correspondence table in Sec.~5.1,
 this operator is nothing but the familiar 
$\xi$ ghost in RNS, carrying picture number 1.
 This is ``inverse'' to $\bar{Q} = \eta_0 + Q_0$, 
 which is the BRST operator appearing at the juncture of connecting 
EPS and RNS. It is related to $Q'_{RNS}$ by the similarity transformation 
 on the RNS side:
\begin{eqnarray}
U Q'_{RNS} U^{-1} =\bar{Q},  \qquad 
\quad 
 U \equiv e^{-X} e^Y e^X e^T \period
\end{eqnarray}
Therefore the counterpart of $\xi$, to be called $\tilde{\xi}$, in 
 the original RNS formalism is given by 
\begin{eqnarray}
\tilde{\xi} = U^{-1} \xi U\period
\end{eqnarray}
We now note that while $X$ and $T$ carry no picture number 
 the operator $Y$ 
 has picture number 1. Therefore, a similarity transformation by $e^{-Y}(\ast)
 e^Y$ 
 is capable of producing an operator with infinitely large positive picture
\footnote{It should nevertheless be stressed that this does not always occur;
 for example, the operator $Q'_{RNS}$ itself is mapped 
 to $\bar{Q}$ with picture number $-1$. Many other 
 examples like this can be constructed.}.  In fact 
 in the case of the operator $\xi$  it is not difficult to prove that the 
transformed $\tilde{\xi}$ does contain
non-vanishing contributions with  arbitrarily large picture number\footnote{
Although we shall not give the technical details, 
the basic reason is that the $e^{3\phi}$ factor present in $Y_2$ when 
 repeatedly applied produce terms with higher and high pictures  together 
 with increasing number of derivatives and product of  
fermionic  fields $b, \eta$ and $\psi^+_I$. A systematic counting
 then shows that the number of derivatives so produced is always sufficient
 to render such product non-vanishing.}.
This is as expected of an operator inverse to $Q'_{RNS}$ and 
 such an operator should be excluded from the Hilbert space 
 by the logic of the RNS side. 
\par\medskip
Thus, we have shown that the cohomology-trivializing operator $\Xi$ 
 can be excluded consistenly in both (E)PS and RNS and  this resolves the
 essential part of the problem.  Admittedly, our argument does not 
 show that at the level of the Hilbert space 
 the notion of ASPC in EPS and that of finite picture in RNS 
are exactly equivalent. In fact it is very unlikely that the image in EPS
 of the space of RNS operators with finite picture
 range  matches precisely with the space of ASPC operators. 
However since the crucial operator $\Xi$ is removed in both spaces and 
 the mapping is one to one (modulo decoupling of the quartets), both 
formulations should have the same non-trivial cohomologies. 

\section{Summary and Discussions}
In this paper, we have succeeded in constructing a similarity 
 transformation which connects, in a well-defined way, the extended 
 version of the pure spinor formalism and the conventional RNS 
 formalism. The BRST charges of these theories are transformed into 
 each other as 
\begin{eqnarray}
&& e^{S_2} e^{\Rtil} \Qhat e^{-\Rtil} e^{-S_2} = \deltatil+\deltabar +
\Qbar \comma \\
&& \Qbar = \eta_0 + Q_0 = 
 e^{-X} e^{Y} e^{X} e^{T} \QpR e^{-T} e^{-X} e^{-Y} e^{X} \comma 
\end{eqnarray}
where the operators in the exponent are fully displayed in appropriate 
sections. We have described the method of construction in some detail 
since this itself is rather powerful and should find applications in 
 other situations as well. When restricted to the proper Hilbert space
 discussed in Sec.~5.2, the mapping provides a direct demonstration of 
 the equivalence of the physical spectrum of these two formulations 
and should prove useful in further investigation of the properties of 
 the EPS and PS formalisms. 

One may have an impression that our similarity transformations look 
rather complicated. Indeed some parts of the calculations required 
a fair amount of effort, due primarily to the $U(5)$ formalism that
 had to be used. This feature, however, is very much expected and 
 unavoidable,  because highly non-trivial 
transmutation of $SO(9,1)$ spinors into vectors must inevitably be involved. 
As is clear from the table summarizing the field-content of 
 EPS and RNS formalisms in Sec.~3.1, only a part of the components 
of space-time Lorentz spinors in EPS are effective in RNS and this splitting 
requires $U(5)$ decomposition. In fact space-time spinors in RNS 
 are realized by spin fields, the components of which are not all 
 independent. In this sense, EPS formalism can be regarded as 
realizing a linearization of the spinor representation in a larger
field space. Deeper understanding of such connections 
 would require a discovery of a universal fundamental action from which 
 one can derive  EPS and RNS formalisms. 

Even with such an underlying action still lacking, by appropriately 
 mapping the firm knowledge available for RNS to the EPS side, 
 one should be able to gain deeper understanding of the properties 
of the EPS and PS formalisms, for example the origin of the rules of 
 computation of the scattering amplitudes, how to 
handle loops, etc. Such a work is underway and 
 we hope to report our findings in a future communication.
\par\bigskip\noindent
{\large\bf Acknowledgment}\par\smallskip\noindent
Y.~K. would like to thank N.~Berkovits 
 for a helpful correspondence. 
The research of Y.K. is supported in part by the 
 Grant-in-Aid for Scientific Research (B) 
No.~12440060 from the Japan  Ministry of Education,  Science
 and Culture. 


\end{document}